\documentclass[twoside]{iopart}
\usepackage[T1]{fontenc}
\usepackage[latin9]{inputenc}
\usepackage{geometry}
\usepackage{color}
\usepackage{amstext}
\usepackage{graphicx}

\makeatletter

\newcommand{\lyxdot}{.}

\usepackage{iopams}
\usepackage{setstack}

\usepackage[numbers, sort&compress]{natbib}

\newcommand{\eqref}[1]{(\ref{#1})}

\makeatother

\begin{document}

\title{Towards an Einstein-Podolsky-Rosen paradox between two macroscopic
atomic ensembles at room temperature}

\author{Q. Y. He$^{1,2,\dagger}$ and M. D. Reid$^{2,*}$}

\address{$^{\text{1}}$State Key Laboratory of Mesoscopic Physics, School of
Physics, Peking University, Beijing 100871 China\\
$^{\text{2}}$Centre for Quantum Atom Optics, Swinburne University
of Technology, Melbourne, Australia}

\ead{$^{\dagger}$qiongyihe@pku.edu.cn, $^{*}$mdreid@swin.edu.au }
\begin{abstract}
Experiments have reported the entanglement of two spatially separated
macroscopic atomic ensembles at room temperature {[}Krauter et al,
Phys. Rev. Lett. \textbf{107}, 080503 (2011) and Julsgaard et al,
Nature \textbf{413}, 400 (2001){]}. We show how an Einstein-Podolsky-Rosen
paradox is realizable with this experiment. Our proposed test involves
violation of an inferred Heisenberg uncertainty principle, which is
a sufficient condition for an EPR paradox. This is a stronger condition
than entanglement. It would enable the first definitive confirmation
of quantum EPR correlations between two macroscopic objects at room
temperature. This is a necessary intermediate step towards a nonlocal
experiment with causal measurement separations. As well as having
fundamental significance, this could provide a resource for novel
applications in quantum technology. 
\end{abstract}

\submitto{\NJP }

\maketitle

\section{Introduction}

The Einstein-Podolsky-Rosen (EPR) paradox was presented in 1935 as
an argument for the incompleteness of quantum mechanics \cite{epr}.
From a modern perspective, this paradox reveals the inconsistency
between local realism and the completeness of quantum mechanics. Based
on the assumption of local realism, the argument was the first clear
illustration of the nonlocality associated with an entangled state.
This has motivated numerous fundamental studies as well as potential
applications in quantum information \cite{wu epr,bell,clauser,grangierepraspect ,bellexp,exp(i)}.
In the original proposal, the quantum state comprised two spatially
separated particles with perfectly correlated positions and momenta
\cite{epr}. It is now understood that the demonstration does not
require such perfect states, but can be inferred from correlations
that violate an inferred Heisenberg inequality - thus signifying this
paradoxical behaviour \cite{eprr,rrmp}. 

Evidence for the EPR paradox has emerged in numerous experiments,
including not only the correlations of single photon pairs, \cite{wu epr,grangierepraspect ,singlephotonepr},
but the amplitudes \cite{exp(i),rrmp,cv_exp,four_wave_mixing,steerCV exp,wagner science,lights_epr}
and Stokes polarization observables \cite{spinpolepr_exp(vi)-1} of
twin optical beams and pulses with macroscopic particle numbers. By
contrast, Bell inequality experiments that test local realism are
affected by the well-known efficiency loophole, and are limited to
microscopic systems \cite{bellexp}. More recently, EPR experiments
\cite{steerCV exp} have been motivated by the work of Wiseman and
co-workers \cite{Wiseman,jonesteerpra,cavaleprsteerineq}. These workers
established that the EPR paradox is a realisation of ``quantum steering'',
a form of nonlocality identified by Schrodinger \cite{Schr=0000F6dinger}
whereby the measurements made by one observer at a location $A$ can
apparently ``steer'' the state of another observer, at a different
location $B$ \cite{steerphotons,eprsteerloss,smtuheprsteer,steer z}.

The task of measuring an EPR paradox between two spatially separated
massive objects - as opposed to massless photons - remains an important
challenge in physics. Despite the significant experimental advances
in the detection of multi-particle entanglement \cite{wineions,blattions},
evidence for an EPR paradox with massive particles is almost nonexistent.
Strong EPR-type entanglement has been verified through the violation
of Bell inequalities for ions \cite{ions} and\textcolor{red}{{} }Josephson
phase qubits \cite{bellsuperconqubts}, but not for significant spatial
separations. Entanglement has also been investigated for cold atom
and BEC systems \cite{heidelepr,bucker,bucker2,estev gross heidel},
neutrons \cite{bea neutrons} and between separated groups of atoms
\cite{ch ens,ch ens-1,cha en,coldatomelis,matkuz ens,shi}. These
observations do not connect directly with an EPR paradox, although
proposals exist \cite{epr_matterwave,beceprolsen,heidelepr,bargillprl,theory_epr_bec,dynamical,becsteer,karen}.
In summary, previous experimental work has yet to detect an EPR paradox
in the correlated macroscopic observables of two massive systems \cite{macro beat,vedral nature}.

Historically, a controversy exists as to the existence of EPR correlations
at large space-like separations. Furry suggested that quantum correlations
could decay with increasing distance between the two systems \cite{pr furry},
thus removing the paradox for large distances. While this can now
be ruled out for massless photons, the situation is not clear for
\emph{massive} objects. Tests of Furry's hypothesis so far have focused
only on the nonlocality between microscopic systems, each of one massless
particle \cite{beautrix furry}. Could Furry's hypothesis possibly
be mass-dependent? Given the use of quantum models to explain the
early universe, and known problems in unifying quantum mechanics with
gravity, this question has an ever-increasing importance in current
physics. 

For macroscopic objects at room temperature, the observation of an
EPR paradox is even more interesting. Here, the question is whether
it is possible to confirm an entanglement involving superpositions
of states that are macroscopically separated in \emph{state} space
for at least one of the sites. The existence of such superpositions
could enhance our understanding of the issues raised in the Schrodinger's
cat paradox \cite{Schr=0000F6dinger-cat}. It is unknown whether an
EPR paradox can be observed for large objects at room temperature.
Our usual intuition is that strong quantum effects of this type will
be masked by thermal motion. The fundamental question is whether macroscopic
quantum paradoxes can be realized at all, or are simply made difficult
by the known physics of decoherence \cite{leggdecoh}. 

In view of these outstanding questions, it is important to analyse
the possibility of realizing an EPR paradox between two systems with
a macroscopic mass. These effects may also have applications in high-precision
interferometry, metrology, cryptography and for reduced quantum uncertainty
for measurements in the presence of quantum memory \cite{REidCrypto,crypsteer,rrmp,berta-1,ralph,cryrr,crychsi,philcry}.
Polzik and co-workers have made a pioneering step in the direction
of realising an EPR paradox between massive objects, in work that
experimentally confirmed the entanglement of two macroscopic ensembles
of gaseous atoms at room temperature \cite{Julsgaard,Krauter,polMuschik,polMuschik2}.
The entanglement is signified by the correlation of position and momenta-like
observables, thus giving a close analogy to the original paradox.
The observables are macroscopic, in that the outcomes are for collective
atomic spins, measured as a very large atom number difference. 
Similar experiments of Lee et al have succeeded in entangling room
temperature macroscopic spins in diamond; however, in that work EPR
observables are not identified \cite{sciencediamond}. 

In this paper, we examine the gas ensemble experiments of Julsgaard
et al \cite{Julsgaard} and Krauter et al \cite{Krauter}. We show
that the reported correlation is not yet strong enough to violate
an inferred Heisenberg inequality. However, our detailed analysis
reveals that this is achievable within the limits of the parameters
reported for the experiments, provided the appropriate conditional
variances are measured. This is the first prediction of an EPR paradox
for room temperature atoms, using a model that accounts for thermal
effects. We also explain how the measurement scheme must be modified
to enable the local measurement of the relevant observables, and 
conclude with a discussion of macroscopic EPR experiments.

\section{Entangling two macroscopic atomic ensembles }

The experiments of Julsgaard, Krauter, Muschik et al \cite{Julsgaard,Krauter,polMuschik}
achieve entanglement of two macroscopic spatially separated atomic
ensembles. The ensembles become entangled when an ``entangling''
light pulse propagates successively through the two ensembles. The
method is based on the proposal of Duan et al \cite{DCZP}. 

We first summarize the theory developed by Duan et al and Julsgaard,
Muschik et al. We will expand on that theory to give a prediction
for the EPR paradox. Let us denote the two spatially separated atomic
ensembles by $A$ and $B$. Schwinger collective spins, $J_{A/B}^{X}$,
$J_{A/B}^{Y}$, $J_{A/B}^{Z}$ are defined for each atomic ensemble,
assumed to contain $N$ atoms. The operators are defined with respect
to two atomic levels which are denoted, for the $j$-th atom in ensemble
$A$, as $|1\rangle_{j}$ and $|2\rangle_{j}$. 

Thus, we define: 
\begin{eqnarray}
J_{A}^{X} & = & \sum_{j=1}^{N}\left(|1\rangle\langle2|_{j}+|2\rangle\langle1|_{j}\right)/2\nonumber \\
J_{A}^{Y} & = & \sum_{j=1}^{N}\left(|1\rangle\langle2|_{j}-|2\rangle\langle1|_{j}\right)/2i\nonumber \\
J_{A}^{Z} & = & \sum_{j=1}^{N}(|1\rangle\langle1|_{j}-|2\rangle\langle2|_{j})/2\label{eq:Define-J-operators}
\end{eqnarray}
 Similar operators $J_{B}^{i}$ are defined for two selected levels
of the ensemble $B$. Each atomic ensemble is prepared initially,
using a detuned pump laser pulse, in an atomic spin coherent state
with a large mean spin $J_{A/B}^{X}$, so that $\langle J_{A/B}^{X}\rangle\sim N$.
This implies the atoms are prepared in a superposition of states $|1\rangle$
and $|2\rangle$. The mean spins are equal and opposite for the ensembles
$A$ and $B$, i.e. $\langle J_{A}^{X}\rangle=-\langle J_{B}^{X}\rangle$. 

In order to observe an EPR paradox, the two ensembles must become
entangled. In the experiments, entanglement is achieved via a detuned
polarized laser pulse, which is called the ``entangling pulse''.
This laser field is described by another set of spin operators, called
the Stokes operators $S^{X}$, $S^{Y}$, $S^{Z}$. In physical terms,
$S^{X}$ is the difference between photon numbers in the orthogonal
$X$ and $Y$ linear polarization directions; $S^{Y}$ is the difference
in number for polarization modes rotated by $\pm\pi/4$; and $S^{Z}$
is the difference between the photon numbers in the two circular polarized
modes defined relative to the propagation direction $Z$. 

These Stokes operators are written in terms of $a_{1,2}$, the operators
for the circular polarized modes, using the Schwinger representation
method. Thus,
\begin{eqnarray}
S^{X} & = & \left(a_{2}^{\dagger}a_{1}+a_{1}a_{2}^{\dagger}\right)/2,\nonumber \\
S^{Y} & = & \left(a_{2}^{\dagger}a_{1}-a_{2}a_{1}^{\dagger}\right)/2i,\nonumber \\
S^{Z} & = & \left(a_{2}^{\dagger}a_{2}-a_{1}^{\dagger}a_{1}\right)/2.\label{eq:schw}
\end{eqnarray}
The light is polarized along the $x$ direction, so that $\langle S^{X}\rangle\sim N_{p}$
where $N_{p}$ is the number of photons in the optical pulse \cite{Julsgaard}. 

Before proceeding, we examine what strength of entanglement will be
required to realize an EPR paradox. EPR's original argument considered
the positions and momentum of each of two particles, in an ideal state
with perfect correlations. However, since this requires infinite momentum
uncertainty for each particle - and therefore infinite energy - it
cannot be achieved physically.

\subsection{EPR paradox inequality}

More practically, the required paradox \emph{can} be constructed with
finite energy, by using two noncommuting observables that are defined
for the systems at each location $A$ and $B$, with finite correlations
stronger than a critical value. These observables, which are analogous
to position and momentum, will be $J_{A}^{Z}$ and $J_{A}^{Y}$ at
$A$, and $J_{B}^{Z}$ and $J_{B}^{Y}$ at $B$. The Heisenberg uncertainty
relation is $\Delta^{2}J_{A}^{Z}\Delta^{2}J_{A}^{Y}\geq\frac{1}{4}|\langle J_{A}^{X}\rangle|^{2}$,
where $\Delta^{2}J_{A}^{Z}$ is the variance for measurement of $J_{A}^{Z}$.
As shown by Reid \cite{eprr}, an EPR paradox is observed when a measurement
at $B$ collapses the wave-function at $A$, to such an extent that
there is a violation of the inferred uncertainty relation. 

The paradox is obtained when the uncertainty product of the conditional
or ``inference'' variances is \emph{less} than a critical value,
i.e., \cite{eprr,caval reid uncerspin,eprspinbach}
\begin{equation}
\Delta_{inf}^{2}J_{A}^{Z}\Delta_{inf}^{2}J_{A}^{Y}<\frac{1}{4}|\langle J_{A}^{X}\rangle|^{2}\,.\label{eq:schepr}
\end{equation}
Here $\Delta_{inf}^{2}J_{A}^{Z}$ is the variance of the conditional
distribution for a result of measurement $J_{A}^{Z}$, given a measurement
made on ensemble $B$, and $\Delta_{inf}^{2}J_{A}^{Y}$ is defined
similarly. The uncertainty $\Delta_{inf}J_{A}^{Z}$ can be viewed
as the average error in the prediction of the result of $J_{A}^{Z}$
given a result for the measurement on ensemble $B$. A useful strategy
is to estimate the result as $g_{Z}j_{B}^{Z}$ where $g_{Z}$ is a
real constant and $j_{B}^{Z}$ is the outcome of measurement $J_{B}^{Z}$.
Similarly, the result for measurement $J_{A}^{Y}$ can be predicted
as $g_{Y}j_{B}^{Y}$, where $g_{Y}$ is a real constant and $j_{B}^{Y}$
is the result of the measurement of $J_{B}^{Y}$. 

Then, we can achieve an EPR paradox if (\ref{eq:schepr}) holds, where
$\Delta_{inf}^{2}J_{A}^{Z}=\Delta^{2}(J_{A}^{Z}-g_{Z}J_{B}^{Z})$
and $\Delta_{inf}^{2}J_{A}^{Y}=\Delta^{2}(J_{A}^{Y}-g_{Y}J_{B}^{Y})$
. Combining these equations, we obtain a gain-dependent criterion
involving actual measured variances: 
\begin{equation}
\Delta^{2}(J_{A}^{Z}-g_{Z}J_{B}^{Z})\Delta^{2}(J_{A}^{Y}-g_{Y}J_{B}^{Y})<\frac{1}{4}|\langle J_{A}^{X}\rangle|^{2}.\label{eq:schepr-1}
\end{equation}
For Gaussian distributions, the inference variances $\Delta^{2}(J_{A}^{Z}-gJ_{B}^{Z})$
and $\Delta^{2}(J_{A}^{Y}-g_{Y}J_{B}^{Y})$ become the variances of
the conditional distributions, for the optimal choice of $g_{X}$
and $g_{Y}$ that will minimize $\Delta^{2}(J_{A}^{Z}-gJ_{B}^{Z})$
and $\Delta^{2}(J_{A}^{Y}-g_{Y}J_{B}^{Y})$ \cite{rrmp}. We mention
here that EPR paradox criteria based on entropic uncertainty relations
have more recently been derived by Walborn et al \cite{entropy},
and become useful in the non-Gaussian case.

We next wish to show that the atomic ensemble experiments can in principle
enable a realisation of the EPR paradox. In order to do this, it is
helpful to first summarize the method used in the experiments in greater
detail. Following the theory of Duan et al \cite{DCZP}, when the
detuned ``entangling'' pulse propagates through the first atomic
ensemble, and the outputs are given in terms of the inputs according
to
\begin{eqnarray}
S_{out}^{Y} & = & S_{in}^{Y}+\alpha J_{in}^{Z},\nonumber \\
S_{out}^{Z} & = & S_{in}^{Z},\nonumber \\
J_{out}^{Y} & = & J_{in}^{Y}+\beta S_{in}^{Z},\nonumber \\
J_{out}^{Z} & = & J_{in}^{Z}.\label{eq:eqnarray1}
\end{eqnarray}
The $\alpha$ and $\beta$ are constants, and the subscripts $out$
and $in$ denote, for the field, the outputs and inputs to the ensemble,
and, for the ensemble, the initial and final states. The $\beta$
is reversed in sign for the second ensemble, so that after successive
interaction with both ensembles, the final output is given in terms
of the first input as 
\begin{eqnarray}
S_{out}^{Y} & = & S_{in}^{Y}+\alpha(J_{in,A}^{Z}+J_{in,B}^{Z}),\nonumber \\
S_{out}^{Z} & = & S_{in}^{Z},\nonumber \\
J_{out,A}^{Y}+J_{out,B}^{Y} & = & J_{in,A}^{Y}+J_{in,B}^{Y},\nonumber \\
J_{out}^{Z} & = & J_{in}^{Z}.\label{eq:eqnarray2}
\end{eqnarray}
Here the subscripts $A$, $B$ denote the spin operators for the $A$,
$B$ ensemble. 

As described in the the original theory of these experiments\cite{Julsgaard,DCZP},
provided $\alpha$ is large enough, the field output $S_{out}^{Y}$
gives a measure of the collective spin $J_{A}^{Z}+J_{B}^{Z}$. This
is a constant of the motion, and is not affected by the interaction.
The Stokes parameter $S_{out}^{Y}$ of the output field is measured
using polarizing beam splitters (PBS) and detectors. As a result,
the atomic ensembles are prepared in a quantum state for which (ideally)
$ $the value of $J_{A}^{Z}+J_{B}^{Z}$ is known and constant. In
practice, technical noise in the preparation implies a state with
reduced noise level in $J_{A}^{Z}+J_{B}^{Z}$. The collective spin
$J_{A}^{Y}+J_{B}^{Y}$ is also a constant of the motion, and hence
the ensembles can be prepared via a second pulse (and by rotating
the atomic spin) in a state with reduced fluctuation in $J_{A}^{Y}+J_{B}^{Y}$.

\subsection{Entanglement and EPR inequalities}

One only has to show that there are reduced fluctuation in these noise
levels according to the uncertainty sum criterion, 
\begin{eqnarray}
\Delta_{ent}=\frac{\Delta^{2}(J_{A}^{Z}+J_{B}^{Z})+\Delta^{2}(J_{A}^{Y}+J_{B}^{Y})}{\bigl|\langle J_{A}^{X}\rangle\bigr|+\bigl|\langle J_{B}^{X}\rangle\bigr|} & < & 1\label{eq:d}
\end{eqnarray}
to signify entanglement between the two atomic ensembles \cite{duan,raymduan}.
Details of how the $\Delta_{ent}$ and the mean spin $\langle J_{A/B}^{X}\rangle$
are measured are given in the Refs. \cite{Julsgaard,DCZP}. For Gaussian,
symmetric systems where the moments of ensemble $A$ and ensemble
$B$ are equal, the entanglement criterion is necessary and sufficient
for entanglement in two-mode systems \cite{duan}. 

The key point, from the perspective of this paper, is that the field-atom
solutions Eq. (\ref{eq:eqnarray2}) give no lower bound, in principle,
to the amount of noise reduction in $J_{A}^{Z}+J_{B}^{Z}$ and $J_{A}^{Y}+J_{B}^{Y}$
that is possible. We comment that ultimately the amount of entanglement
attainable will be limited by the uncertainty relation and the finite
nature of the Schwinger spins \cite{soremolme,cj-1}, but that, while
significant in some BEC proposals \cite{becsteer,bargillprl,heidelepr},
this limitation becomes unimportant here, because of the large numbers
of atoms involved. Thus, a physical regime exists for which
\begin{equation}
\Delta^{2}(J_{A}^{Z}+J_{B}^{Z})\rightarrow0,\,\,\,\,\Delta^{2}(J_{A}^{Y}+J_{B}^{Y})\rightarrow0.\label{eq:limit}
\end{equation}
 In that case, the measurement of $J_{B}^{Z}$ will imply the result
for the measurement of $J_{A}^{Z}$, with \emph{no} uncertainty. Similarly,
the result for $J_{B}^{Y}$ will imply, precisely, the result for
$J_{A}^{Z}$. Therefore the variances of the conditional distributions
$P(J_{A}^{Z}|J_{B}^{Z})$ and $P(J_{A}^{Y}|J_{B}^{Z})$ become zero,
and the EPR condition (\ref{eq:schepr}) will be satisfied, with $g_{Z}=g_{X}=1$.
The solutions Eq. (\ref{eq:eqnarray1}) on which the prediction is
based are valid in the limit where damping effects can be neglected.
Duan et al analyzed the full atomic solutions, and reported that this
regime is achievable when $N\sim N_{p}$ provided the field detunings
are much greater than spontaneous emission rates \cite{DCZP}. 

For realistic systems, the variances $\Delta^{2}(J_{A}^{Z}+J_{B}^{Z})$
and $\Delta^{2}(J_{A}^{Y}+J_{B}^{Y})$ will not, however, be zero.
The measure of entanglement given by $\Delta_{ent}$ can indicate
the strength of correlation that is needed for an EPR paradox, when
\begin{equation}
\Delta_{ent}<0.5.\label{eq:dhalf}
\end{equation}
for symmetric ensembles. This follows because $\Delta_{ent}<0.5$
implies $\Delta^{2}(J_{A}^{Z}+J_{B}^{Z})+\Delta^{2}(J_{A}^{Y}+J_{B}^{Y})<(\bigl|\langle J_{A}^{X}\rangle\bigr|+\bigl|\langle J_{B}^{X}\rangle\bigr|)/2$,
which in turn implies that $\Delta(J_{A}^{Z}+J_{B}^{Z})\Delta(J_{A}^{Y}+J_{B}^{Y})<(\bigl|\langle J_{A}^{X}\rangle\bigr|+\bigl|\langle J_{B}^{X}\rangle\bigr|)/4$
is true \cite{rrmp}. If for the experiment, we measure that $\bigl|\langle J_{A}^{X}\rangle\bigr|=\bigl|\langle J_{B}^{X}\rangle\bigr|$,
then the measured variances must also satisfy $\Delta(J_{A}^{Z}+J_{B}^{Z})\Delta(J_{A}^{Y}+J_{B}^{Y})<\bigl|\langle J_{A}^{X}\rangle\bigr|/2$.
This satisfies the condition (\ref{eq:schepr-1}), with the choice
of $g_{Z}=g_{Y}=1$. In general, it has been shown for Gaussian states
(where losses and thermal noise are included) that the condition (\ref{eq:schepr})
for realizing the EPR paradox is more difficult to achieve than (\ref{eq:d}),
that for entanglement. 

We learn from this Section that there are two important issues to
be considered in realizing the EPR paradox. The first is that greater
correlations are required for the EPR paradox than to realize simple
entanglement. The EPR experiment will therefore be more sensitive
to decoherence effects. We address this first issue in Section 4,
by analyzing in detail the theoretical model presented by Muschik
et al \cite{polMuschik}, that accounts for such effects in a recent
experiment. Second, the EPR condition (\ref{eq:schepr}) also specifies
that the measurements of the four observables $J_{A}^{Z}$, $J_{B,}^{Z}$,
$J_{A}^{Y}$ and $J_{B}^{Y}$ be made \emph{locally}. We address this
issue in the next Section.

\section{Measurement of the EPR paradox}

\begin{figure}
\begin{centering}
\includegraphics[width=0.7\columnwidth]{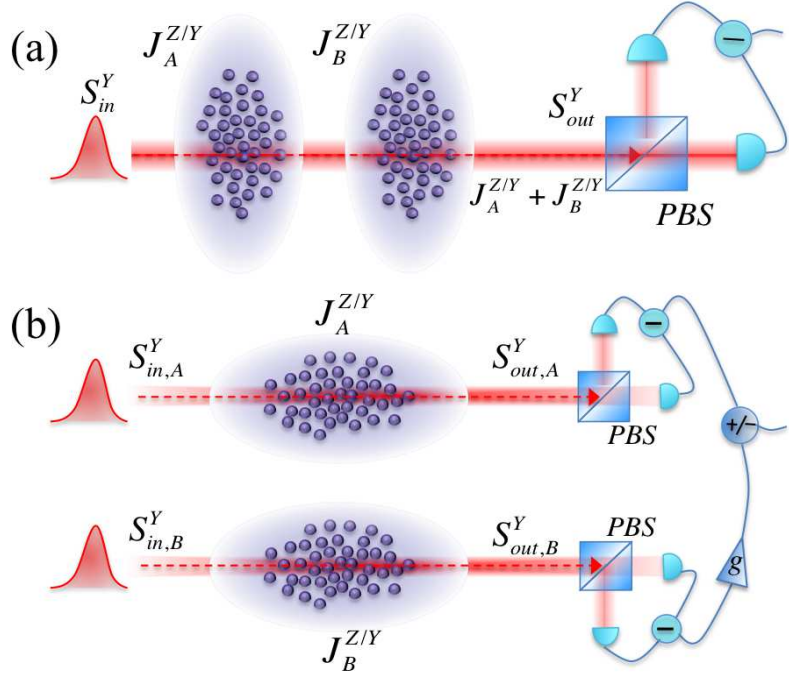}
\par\end{centering}

\caption{\textcolor{black}{Schematic diagram of how to measure an EPR paradox
between two atomic ensembles. (a) The arrangement to measure the entanglement
of the two ensembles, as in the experiments \cite{Julsgaard,Krauter,polMuschik}.
The variance of the collective spin $J_{A}^{Z}+J_{B}^{Z}$ (or $J_{A}^{Y}+J_{B}^{Y}$)
is measured by passing a ``verifying'' light pulse (with input Stokes
operator $S_{in}^{Y}$) through both ensembles. Detection of the output
pulse Stokes operator ($S_{out}^{Y}$) allows measurement of either
$J_{A}^{Z}+J_{B}^{Z}$ or $J_{A}^{Y}+J_{B}^{Y}$, depending on the
selection of rotation of the atomic spin of the ensemble. (b) An experiment
to demonstrate the EPR paradox between two atomic ensembles. Independent
local measurements are made on the atomic ensembles $A$ and $B$,
by utilizing two verifying pulses. The choice of whether to measure
$J^{Z}$ or $J^{Y}$ at each ensemble is made independently and after
preparation of the EPR state (which is done by the entangling pulses).
The variable gain factor $g$ is selected to enable a final optimal
measurement of the conditional variance $\Delta^{2}(J_{A}^{Z}+g_{Z}J_{B}^{Z})$
(or $\Delta^{2}(J_{A}^{Y}+g_{Y}J_{B}^{Y})$). \label{fig:Schematic-atomic_steering}}}
\end{figure}

The experiments \cite{Julsgaard,Krauter} detect entanglement using
a ``verifying pulse'' that propagates through the two ensembles
(Fig. \ref{fig:Schematic-atomic_steering}a). The Stokes observable
$S_{out}^{Y}$ of the transmitted pulse is measured via the polarizing
beam splitter. Using equations (\ref{eq:eqnarray2}), the outcome
for the Stokes observable allows the value of the collective spin
$(J_{B}^{Z}+J_{A}^{Z})$ to be inferred. The measurement of $(J_{B}^{Y}+J_{A}^{Y})$
proceeds similarly, after rotation of the atomic spin. The measurement
scheme thus establishes the \emph{collective} spin by a final projective
measurement at one location. The quality of the measurement improves
as $\alpha$ becomes larger. 

For smaller $\alpha$, the noise levels of the output become dominated
by the vacuum noise $\Delta^{2}S_{in}^{Y}$ of the input pulse. In
that case, quantum noise squeezing of the input Stokes observable
would improve the signal to noise ratio of the measurement.

\subsection{Modification of the measurement strategy }

To carry out an EPR experiment, it is necessary to make \emph{local}
measurements of $J_{B}^{Z}$ and $J_{A}^{Z}$ individually, and to
then obtain the collective spin sum $(J_{B}^{Z}+J_{A}^{Z})$ by addition
of the measurement outcomes. Similarly, one should measure $(J_{B}^{Y}+J_{A}^{Y})$
in a separate experimental procedure.

In order to achieve these goals, one possible strategy is to use two
verifying pulses, defined with Stokes parameters $S_{A}^{Y}$ and
$S_{B}^{Y}$, propagating through each ensemble $A$ and $B$ respectively
(as shown in Fig. \ref{fig:Schematic-atomic_steering}b). The outputs
in terms of the inputs are given by the solutions Eq. (\ref{eq:eqnarray1}),
so that
\begin{eqnarray}
S_{out,A}^{Y} & = & S_{in,A}^{Y}+\alpha J_{in,A}^{Z},\nonumber \\
S_{out,B}^{Y} & = & S_{in,B}^{Y}+\alpha J_{in,B}^{Z}.\label{eq:measure}
\end{eqnarray}
Measurement of $S_{out,A}^{Y}$ at one location, and $S_{out,B}^{Y}$
at the other location, enables local determination of $J_{A}^{Z}$
and $J_{B}^{Z}$, as required for a test of EPR nonlocality. The local
measurements of $J_{A}^{Y}$ and $J_{B}^{Y}$ can be made similarly.
The quality of the measurement of the atomic spins $J_{A/B}^{Z}$
improves when $\alpha$ is large, or if the input fields are ``squeezed'',
so that $\Delta S_{in,A}^{Y}\rightarrow0$ and $\Delta S_{in,B}^{Y}\rightarrow0$.

\section{Detailed calculation for an engineered dissipative system}

The recent experiment of Krauter et al \cite{Krauter} employs an
engineered dissipative process to generate a long-lived entanglement
between the atomic ensembles. The engineered dissipation approach
offers a means to tailor dissipative processes, to enhance the generation
of entangled or EPR states \cite{dynamical,eng nonlinear dis}. Entanglement
values of $\Delta_{ent}\sim0.9$ were achieved, using this process,
in reasonable agreement with the theory presented by Muschik, Polzik
and Cirac (MPC) \cite{polMuschik,polMuschik2} for this experiment.
We therefore analyse that theory, to calculate whether the EPR paradox
is predicted for a complete and realistic atomic model.

The MPC model introduces operators $A=(\mu J_{A}^{-}-vJ_{B}^{+})/\sqrt{N}$,
$B=(\mu J_{B}^{-}-vJ_{A}^{+})/\sqrt{N}$, where $J_{A/B}^{\pm}$ denote
collective spin operators with $J_{A/B}^{-}=\sum_{j=1}^{N}|\uparrow\rangle\langle\downarrow|_{j}$
and $J_{A/B}^{+}=\sum_{j=1}^{N}|\downarrow\rangle\langle\uparrow|_{j}$
such that $J_{A/B}^{Y}=\left(J_{A/B}^{+}+J_{A/B}^{-}\right)/2$ and
$J_{A/B}^{Z}=i\left(J_{A/B}^{+}-J_{A/B}^{-}\right)/2$, and $\mu^{2}-\nu^{2}=1$.
Here, the two-level states $|\uparrow\rangle,$$|\downarrow\rangle$
given in the definitions for the spins of ensemble $A$/ $B$ refer
only to the ensemble $A$/$B$ respectively. The $\mu$ and $\nu$
characterize a squeezed state with squeeze parameter $r$, where $\mu=\cosh\left(r\right)$
and $\nu=\sinh\left(r\right)$. The $\mu$ and $\nu$ are functions
of the Zeeman splitting of the two atomic energy levels and of the
detuning of the laser that couples these levels to the excited states
and to vaccum modes, as detailed in Ref. \cite{polMuschik}. The work
of Muschik et al \cite{polMuschik} shows how the correlations are
described by a master equation \cite{Krauter}
\begin{eqnarray}
\frac{d\rho}{dt} & = & \frac{d\Gamma}{2}\times(A\rho A^{\dagger}-A^{\dagger}A\rho+B\rho B^{\dagger}-B^{\dagger}B\rho+H.c.)\nonumber \\
 &  & \,\,\,\,\,\,\,\,\,\,\,+\mathcal{L}_{noise}\rho\label{eq:master}
\end{eqnarray}
where $\rho$ is the atomic density operator, $d$ is the optical
depth of an ensemble and $\Gamma$ is the single atom radiative decay.
The $\mathcal{L}_{noise}\rho$ represent the detrimental processes
such as single atom spontaneous emission noise, thermal effects and
collisions which counter the development of the entangled state, but
which are included in the analysis to give a realsitic prediction.
The Lindblad terms given in the parentheses arise from the engineered
dissipative mechanism that drives the system into the EPR state \cite{Krauter}.
For large optical depth $d$, these entangling effects are enhanced.
Other parameters are the number of atoms $N_{\uparrow}$ and $N_{\downarrow}$
in each of the two levels; and the normalized population $P_{2}(t)=(N_{\uparrow}-N_{\downarrow})/N_{2}(t)$
where $N_{2}=N_{\uparrow}+N_{\downarrow}$ is the number of atoms
in the two-level system.

\subsection{Correlation dynamics}

\begin{figure}
\begin{centering}
\includegraphics[width=0.7\columnwidth]{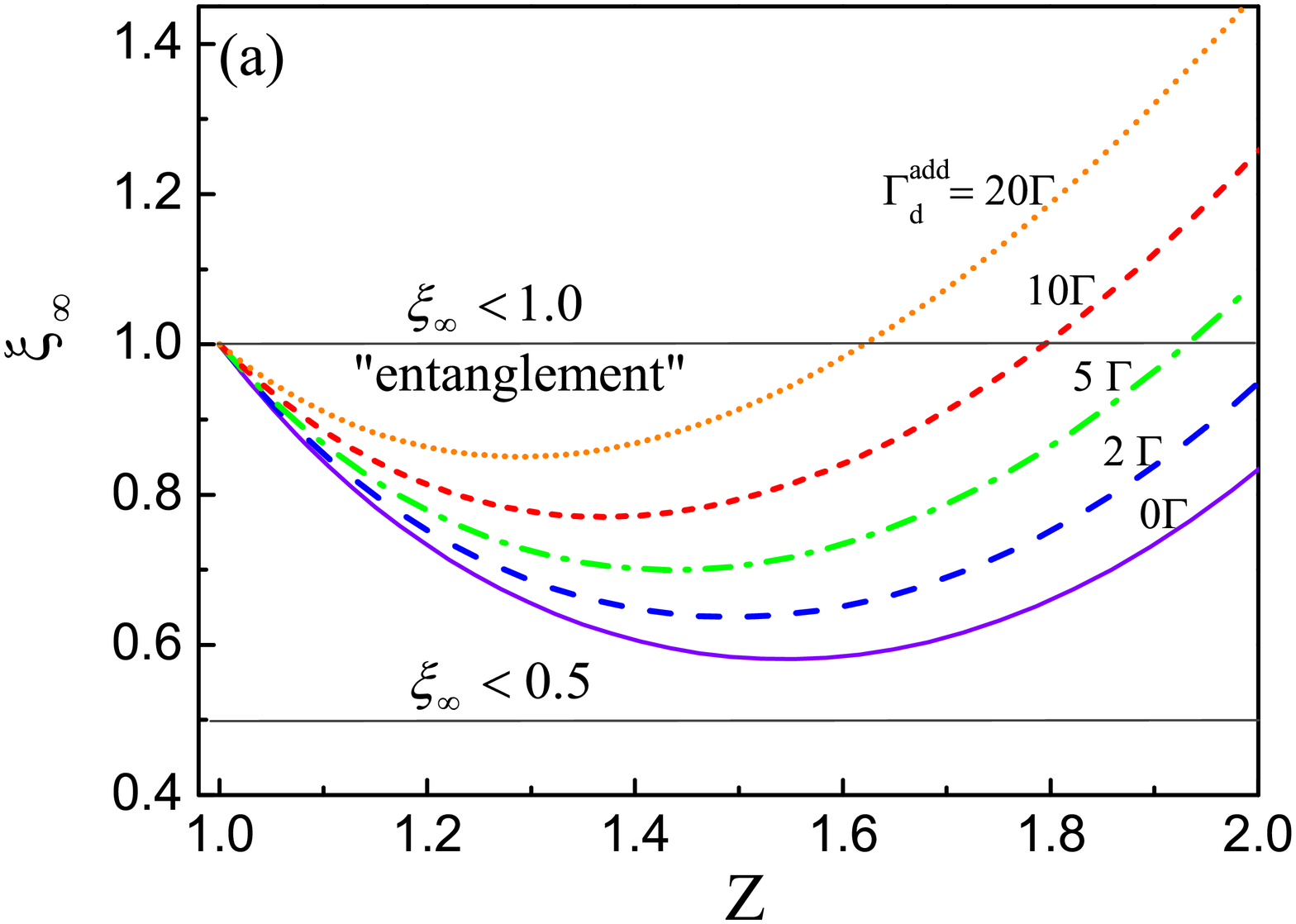}
\par\end{centering}

\begin{centering}
\includegraphics[width=0.7\columnwidth]{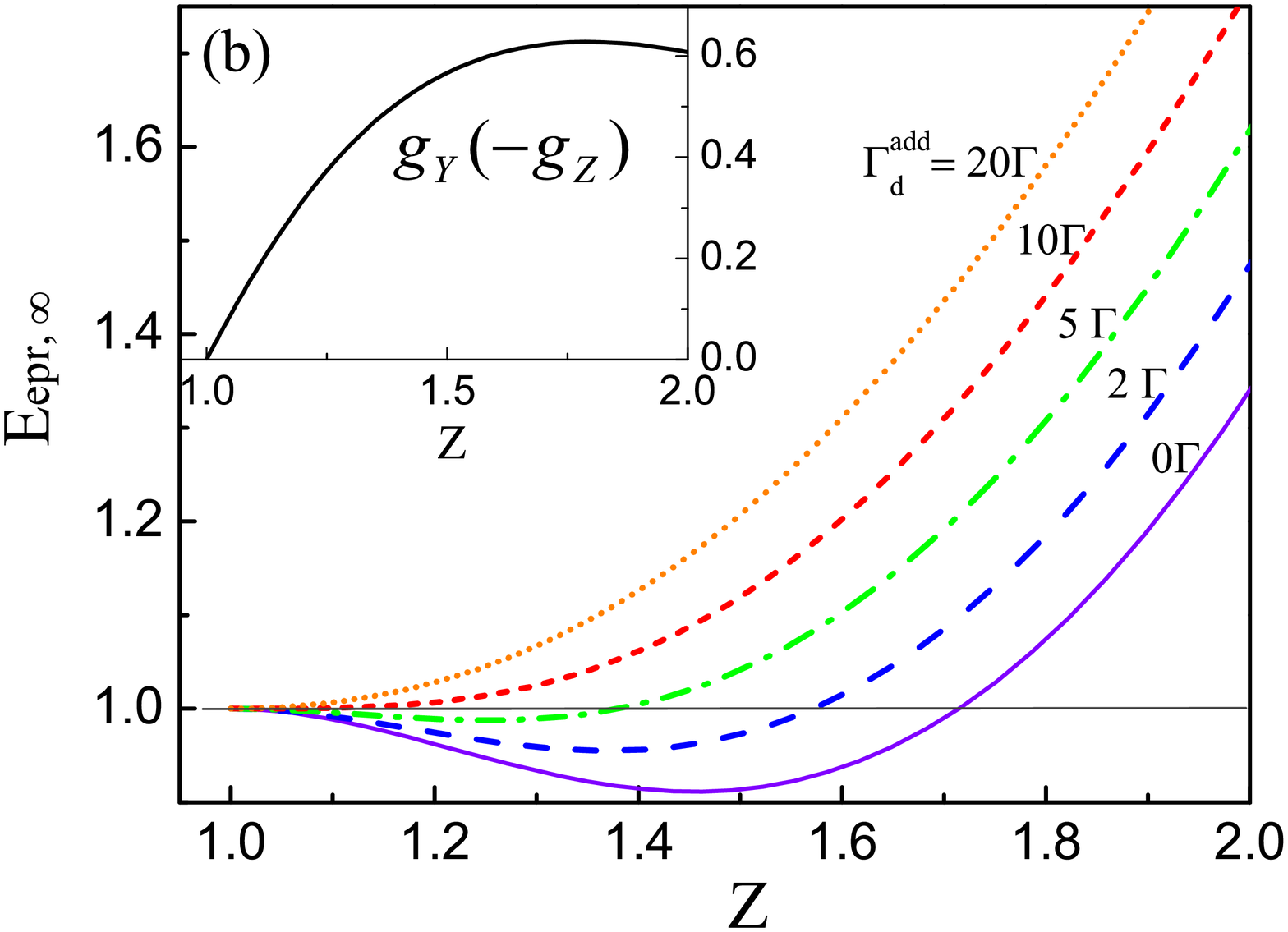}
\par\end{centering}

\centering{}\caption{\textcolor{black}{(Color online) Realizing a stable steady-state EPR
paradox using the engineered dissipatively driven system of Muschik
et al \cite{polMuschik}. (a) The predictions for the steady-state
entanglement $\Delta_{ent}=\xi_{\infty}$ versus $Z=(\mu-v)^{-1}$
for an optical depth $d=30$ per ensemble. We use the parameters presented
in Ref. \cite{polMuschik}. $\xi_{\infty}<1$ indicates entanglement
of the two ensembles, and the observation of $\Delta_{ent}=\xi_{\infty}<0.5$
would imply the correlations of an EPR paradox. (b) The predictions
for the EPR paradox parameter (\ref{eq:Esteer_infinite}) are given
using the same parameters as in (a), but with optimal values of gain
factor $g_{Y}$ and $g_{Z}$. $E_{epr,\infty}<1$ indicates an EPR
paradox. The inset shows the optimal gain factor $g_{Y}(-g_{Z})$
used to minimize $E_{epr,\infty}$. Here, $\Gamma_{cool}=\mu^{2}\Gamma$
and $\Gamma_{heat}=v^{2}\Gamma$. The dephasing rate $\Gamma_{d}=\Gamma_{d}^{rad}+\Gamma_{d}^{add}$
consists of a radiative part $\Gamma_{d}^{rad}=2(\mu^{2}+v^{2})\Gamma$
which is due to light-induced transitions, and an additional term
$\Gamma_{d}^{add}$ which summarizes all nonradiative sources of dephasing.
The violet line is for pure radiative damping, $\Gamma_{d}^{add}=0$.
$\Gamma$ is the single particle decay rate}.\label{fig:atomic steering}\textcolor{red}{{}
}}
\end{figure}

MPC derive the following dynamical equations for the evolution of
the atomic spin correlations:
\begin{eqnarray}
\frac{d}{dt}\langle\left(J_{A/B}^{Z}\right)^{2}\rangle & = & -[\tilde{\Gamma}+d\Gamma P_{2}(t)\bigl]\langle\left(J_{A/B}^{Z}\right)^{2}\rangle\nonumber \\
 &  & +\frac{N}{4}\bigl[\tilde{\Gamma}+d\Gamma P_{2}(t)^{2}(\mu^{2}+\nu^{2})\bigr],\nonumber \\
\frac{d}{dt}\langle\left(J_{A/B}^{Y}\right)^{2}\rangle & = & -[\tilde{\Gamma}+d\Gamma P_{2}(t)\bigl]\langle\left(J_{A/B}^{Y}\right)^{2}\rangle\nonumber \\
 &  & +\frac{N}{4}\bigl[\tilde{\Gamma}+d\Gamma P_{2}(t)^{2}(\mu^{2}+\nu^{2})\bigr],\nonumber \\
\frac{d}{dt}\langle J_{A}^{Z}J_{B}^{Z}\rangle & = & -[\tilde{\Gamma}+d\Gamma P_{2}(t)\bigl]\langle J_{A}^{Z}J_{B}^{Z}\rangle\nonumber \\
 &  & +\frac{N}{2}\mu\nu d\Gamma P_{2}(t)^{2},\nonumber \\
\frac{d}{dt}\langle J_{A}^{Y}J_{B}^{Y}\rangle & = & -[\tilde{\Gamma}+d\Gamma P_{2}(t)\bigl]\langle J_{A}^{Y}J_{B}^{Y}\rangle\nonumber \\
 &  & -\frac{N}{2}\mu\nu d\Gamma P_{2}(t)^{2},
\end{eqnarray}
where $\tilde{\Gamma}=\Gamma_{cool}+\Gamma_{heat}+\Gamma_{d}$, $\Gamma_{cool}(\Gamma_{heat})$
is the total single-particle cooling (heating) rate and $\Gamma_{d}$
is the total dephasing rate. The entangling terms of Eq. (\ref{eq:master})
arise as those being proportional to $d$ and will drive the system
into an entangled EPR state. From this set of equations, in order
to evaluate the predictions for the entanglement parameter $\Delta_{ent}$,
MPC derive an equation describing the evolution of the variances $\Delta^{2}(J_{A}^{Y/Z}\pm J_{B}^{Y/Z})$.
The results of solving these equations in the steady-state are given
in Fig (\ref{fig:atomic steering}), in the first panel, which plots\textcolor{black}{{}
the steady-state entanglement $\Delta_{ent}=\xi_{\infty}$}. This
is clearly adequate for demonstrating entanglement, since variance
sums of this type are an entanglement witness, as explained in the
previous sections. These types of correlations were measured in the
original experiment, as illustrated in Fig (\ref{fig:Schematic-atomic_steering}). 

However, as we explain in greater detail below, these variance sums
are\textbf{ not} sufficient to demonstrate the stronger EPR requirement
that we are interested in here. In fact, variance sums \textbf{can}
be used as an EPR witness, provided correlations are strong enough
and causality requirements are satisfied. The problem is that under
the conditions of these experiments, with relatively imperfect correlations,
variance sums are not an efficient witness. In simple terms, the use
of this type of witness does not lead to evidence for an EPR paradox.
The observed correlation strength in these experiments is not strong
enough. 

For the purpose of this paper, it is better, instead, to directly
investigate the conditional variances which are at the heart of the
inferred Heisenberg inequality.

\subsection{Conditional variance calculation}

To obtain the optimal prediction for the EPR paradox we need to modify
the MPC analysis, to calculate the evolution of the \emph{conditional}
variances, $\Delta_{inf}^{2}(J_{A}^{Y})=\Delta^{2}(J_{A}^{Y}\pm g_{Y}J_{B}^{Y})$
and $\Delta_{inf}^{2}(J_{A}^{Z})=\Delta^{2}(J_{A}^{Z}\pm g_{Z}J_{B}^{Z})$,
as given by Eq. (\ref{eq:schepr-1}). These EPR variances can be measured
by the arrangement of Fig. \ref{fig:Schematic-atomic_steering}b.
The equations are
\begin{eqnarray*}
\frac{d}{dt}\Delta_{inf}^{2}(J_{A}^{Y})\Bigr\} & = & -[\tilde{\Gamma}+d\Gamma P_{2}(t)\bigl]\Bigl\{\Delta_{inf}^{2}(J_{A}^{Y})\Bigr\}\\
 &  & +\frac{N(1+g_{Y}^{2})}{4}\bigl[\tilde{\Gamma}+d\Gamma P_{2}(t)^{2}(\mu^{2}+\nu^{2})\bigr]\mp Ng_{Y}\mu\nu d\Gamma P_{2}(t)^{2}\\
\\
\frac{d}{dt}\Delta_{inf}^{2}(J_{A}^{Z})\Bigr\} & = & -[\tilde{\Gamma}+d\Gamma P_{2}(t)\bigl]\Bigl\{\Delta_{inf}^{2}(J_{A}^{Z})\Bigr\}\\
 &  & +\frac{N(1+g_{Z}^{2})}{4}\bigl[\tilde{\Gamma}+d\Gamma P_{2}(t)^{2}(\mu^{2}+\nu^{2})\bigr]\pm Ng_{Z}\mu\nu d\Gamma P_{2}(t)^{2}.
\end{eqnarray*}
The steady state solutions are given by 
\begin{eqnarray*}
\Delta_{inf}^{2}(J_{A}^{Y})_{\infty} & = & \frac{A_{Y}\bigl[\tilde{\Gamma}+d\Gamma P_{2,\infty}^{2}(\mu^{2}+\nu^{2})\bigr]\mp Ng_{Y}\mu\nu d\Gamma P_{2,\infty}^{2}}{[\tilde{\Gamma}+d\Gamma P_{2,\infty}\bigl]},\\
\\
\Delta_{inf}^{2}(J_{A}^{Z})_{\infty} & = & \frac{A_{Z}\bigl[\tilde{\Gamma}+d\Gamma P_{2,\infty}^{2}(\mu^{2}+\nu^{2})\bigr]\pm Ng_{Y}\mu\nu d\Gamma P_{2,\infty}^{2}}{[\tilde{\Gamma}+d\Gamma P_{2,\infty}\bigl]}
\end{eqnarray*}
where $A_{Y}=N(1+g_{Y}^{2})/4$ and $A_{Z}=N(1+g_{Z}^{2})/4$, $P_{2,\infty}=\lim_{t\rightarrow\infty}P_{2}(t)$
and we have chosen the gains such that: 
\begin{equation}
g_{Y}=-g_{Z}=\frac{\pm\mu vd\Gamma P_{2,\infty}^{2}}{\bigl[\tilde{\Gamma}+d\Gamma P_{2,\infty}^{2}(\mu^{2}+\nu^{2})\bigr]/2}\label{eq:gmin}
\end{equation}
to minimize the conditional EPR variances. The steady state solution
for $\bigl|\langle J_{A,B}^{X}\rangle\bigr|$ is given by Muschik
et al \cite{polMuschik}, as 
\begin{eqnarray}
\bigl|\langle J_{A,B}^{X}\rangle\bigr|_{\infty} & = & \frac{N}{2}P_{2,\infty}.
\end{eqnarray}

These inferred variances correspond to quantities that would be measurable
in the second type of experiment illustrated in Fig (\ref{fig:Schematic-atomic_steering}).
So far, this type of experiment with local measurements at each site
has not been carried out. The important issue is to be able to measure
the spin of each atomic ensemble individually, in order to calculate
the inferred or conditional variance.

\subsection{EPR paradox predictions}

We can define from the EPR paradox condition (\ref{eq:schepr}) the
normalized EPR paradox parameter
\begin{equation}
E_{epr}(A|B)=\frac{\Delta_{inf}(J_{A}^{Z})\Delta_{inf}^{2}(J_{A}^{Y})}{\frac{1}{2}\bigl|\langle J_{A}^{X}\rangle\bigr|}\label{eq:epra|B}
\end{equation}
 that gives an indication of the amount of EPR paradox. The EPR paradox
is obtained when $E_{epr}(A|B)<1$ and is strongest as $E_{epr}(A|B)\rightarrow0$
\cite{rrmp}. We note the asymmetry of this definition, with relation
to the subsystems $A$ and $B$. 

An EPR paradox is also obtained when $E_{epr}(B|A)<1$. Either condition
($E_{epr}(A|B)<1$ or $E_{epr}(B|A)<1$) is sufficient to demonstrate
the paradox, and for asymmetric systems, where the parameters are
not equal, this fact can become important and interesting \cite{steer thmurray,onewaysteer,multi_nonlocality,kasym}.
Recent work identifies $E_{epr}(A|B)<1$ as a criterion to verify
an ``EPR steering'' of Alice's system $A$, by Bob's measurements
on system $B$ \cite{Wiseman,jonesteerpra,cavaleprsteerineq}. The
asymmetry of definition is inherent in the original argument of EPR. 

The steady state solution for the EPR parameter is given as 
\begin{equation}
E_{epr,\infty}=\frac{\Delta_{inf}(J_{A}^{Z})_{\infty}\,\,\Delta_{inf}^{2}(J_{A}^{Y}){}_{\infty}}{\frac{1}{2}\bigl|\langle J_{A}^{X}\rangle\bigr|_{\infty}}\label{eq:Esteer_infinite}
\end{equation}
and is plotted in the Figure 2. Entanglement is verified if $\Delta_{g.ent}<1$
where 
\begin{equation}
\Delta_{g,ent}=\xi_{g,\infty}=\frac{\Delta_{inf}^{2}(J_{A}^{Z})_{\infty}+\Delta_{inf}^{2}(J_{A}^{Y})_{\infty}}{\left(\bigl|\langle J_{A}^{X}\rangle\bigr|_{\infty}+\bigl|g_{Y}g_{Z}||\langle J_{B}^{X}\rangle\bigr|_{\infty}\right)/2},\label{eq:ent_g-1}
\end{equation}
and we introduce the notation $\Delta_{g,ent}$ to remind us that
there is a dependence on the experimental parameters $g_{Y}$ and
$g_{Z}$ \cite{entasym,asG,man}. This criterion is more general than
the symmetric criterion (\ref{eq:d}) of Duan et al \cite{duan}.
We note that $\Delta_{ent}=\Delta_{g,ent}$ when $g_{Y}=-g_{Z}=1$.
We introduce the notation $\xi_{\infty}$ used by Muschik et al: $\xi_{g,\infty}$
is the value of $\Delta_{g,ent}$ in the steady state, and $\epsilon_{\infty}$
is the steady state value of $\Delta_{ent}$.

An important point is that the choice for $g_{X}$ and $g_{Y}$ to
minimize $E_{epr}$ is \emph{not} the same choice that will minimize
the entanglement parameter $\Delta_{g,ent}$. In fact, for ensembles
symmetric under interchange $A\longleftrightarrow B$, it is possible
to show that $\Delta_{g,ent}$ is minimized by $g_{Y}=-g_{Z}=1$.
In this case, $\Delta_{g,ent}\equiv\Delta_{ent}$. The steady state
value of the entanglement parameter $\Delta_{ent}$ denoted $\xi_{\infty}$
is plotted in Fig. \ref{fig:atomic steering}a, in agreement with
Muschik et al \cite{polMuschik}. As explained in Section 3, the observation
of $\Delta_{ent}=\xi_{\infty}<0.5$ would imply the correlations of
an EPR paradox. However, we see from the Fig. \ref{fig:atomic steering}a
that this cannot be achieved, for the parameter range chosen. 

We plot the prediction for the EPR paradox in Fig. 2b, using the optimal
arrangement of Fig 1b, where the gain factors $g_{Y}$ and $g_{Z}$
are selected as in (\ref{eq:gmin}) different to $1$. We use the
same parameters as for Figure 2a, that are selected by Muschik et
al \cite{polMuschik} to model the ensembles realistically in the
experimental regime. Values of $E_{epr,\infty}\sim0.9$ are predicted
for mainly pure radiative damping. In the same regime, values of $\Delta_{ent}\sim0.6$
for the steady state entanglement are predicted. The EPR paradox parameter
is more sensitive to dephasing than is entanglement. Given that the
experiment has achieved a steady state entanglement of $\Delta_{ent}\sim0.9$
however, a realisation of an EPR paradox would seem feasible.

\section{Discussion}

The evidence for quantum nonlocality becomes compelling if the entangled
systems are separated to the extent that the measurements made on
the atomic ensembles by the verifying pulses are space-like separated
events \cite{steer z}. This consolidates the locality premise because
then one can rule out a causal influence on the system $A$ due to
measurements at $B$. The measurement events are made in the time
taken for the verifying pulse to interact with the individual ensemble.
We denote this time as $\Delta T$. Based on the description of the
experiment of Ref. \cite{Julsgaard,Krauter}, for that case, $\Delta T\sim0.45$
ms. 

The entangling pulse (or engineered entanglement mechanism, as discussed
above) must in the first instance be able to propagate through both
ensembles to create the entanglement. It is also necessary that the
entanglement lifetime exceed the measurement time, a condition that
is clearly satisfied for the experiments of \cite{Krauter} which
report entanglement times of up to $0.04$s. 

The separation between the ensembles must then be at least $D>c\Delta T$,
which gives a requirement of much larger than room size separations
found in current experiments. We therefore propose that the Polzik-type
set-up, which is spatially separated but not strictly causal, would
be a first step in which one obtains correlations of enough strength
to guarantee the EPR paradox. A necessary subsequent experiment using
pulsed entanglement with causal separation would be needed for a full
EPR paradox demonstration. We note that pulsed local oscillator measurements
provide well-defined properties analogous to the quadrature variables
defined here \cite{DrummondPulsedQuad}. Such quadrature measurements
have been carried out previously for purely optical systems \cite{SlusherGrangier}.
They are not impossible in a future, fully causal experiment on correlated
atomic ensembles.

A second potential criticism of the proposed experiment is the nature
by which the macroscopic atomic spin of the atomic ensembles are measured.
The value for the spin is inferred via the relationship given by equations
(\ref{eq:measure}), based on the measurement of the Stokes field
observables. The validity of the measurement then depends on the correctness
of the quantum based equations, which involves the question of Gaussian
mode-coupling constants \cite{Rosenberger} and the corresponding
noise calibration issues. In this respect, the proposals such as given
in Ref. \cite{heidelepr,epr_matterwave} to realize an EPR paradox
based on the spin observables of two groups of atoms of a Bose Einstein
Condensate (BEC), by using four-wave mixing or molecular dissociation,
provide an important alternative. Here, each group of atoms can be
constrained spatially by the potential well of an optical lattice.
While the spatial separations are therefore limited, the advantage
is that atomic populations are measured directly by atomic imaging,
which simplifies calibration issues. Other advantages that exist include
the possibility of longer decoherence times \cite{eg decoh} and reduced
atomic dephasing which the theory summarized in Section IV indicates
is detrimental to the EPR correlations. 

Perhaps the most fascinating feature of the current proposal is the
macroscopic nature of the EPR correlations. The EPR observables are
the Schwinger atomic spins, which correspond to the difference in
the numbers of atoms populating two specified atomic states. For these
experiments, the total atom numbers are large ($N\sim10^{13}$ atoms).
The atom number differences do not need to be measured microscopically
$-$ that is, with a microscopic precision where single atoms are
distinguished$-$ in order to attain the EPR paradox. The reason for
this is understood by examining the EPR paradox condition (\ref{eq:schepr}),
and noticing that the right side depends on the total mean spin $\langle J_{A}^{X}\rangle,$
which is given as $\langle J^{X}\rangle\sim N$ \cite{noisebell}.
The quantum noise level of the uncertainty relation becomes a macroscopically
measurable quantity, thus enabling the paradox to be realized based
on the ratio (\ref{eq:epra|B}) only.

\section{Conclusion}

In summary, we have examined the possibility of detecting an EPR paradox
between two macroscopic atomic ensembles at room temperature, based
on the experiments that have realized an entanglement between the
ensembles. Although the realisation of an EPR paradox is more difficult
than for entanglement, detailed models that account for decoherence
effects allow prediction of the EPR paradox, for room temperature
atoms. We have shown that the measurement scheme must be modified,
to enable local measurements on each ensemble to be performed. The
proposed experiment would be a convincing demonstration of quantum
nonlocality in the form of the EPR paradox and quantum steering for
truly macroscopic objects.

\ack{}{}

We thank P. D. Drummond for useful discussions and feedback. We acknowledge
support from the Australian Research Council for funding via ACQAO
COE, Discovery, and DECRA grants. Q. Y. H. wishes to thank the support
by the National Natural Science Foundation of China under Grant No.
11121091 and 11274025.

\end{document}